\documentclass[aps,prd,preprint,groupedaddress,amsmath,amssymb]{revtex4}
\usepackage{epsfig}

\begin{document}

\def\d{{\rm d}}
\def\eps{\varepsilon}
\def\lp{\left. }
\def\rp{\right. }
\def\lr{\left( }
\def\rr{\right) }
\def\le{\left[ }
\def\re{\right] }
\def\lg{\left\{ }
\def\rg{\right\} }
\def\lb{\left| }
\def\rb{\right| }
\def\beq{\begin{equation}}
\def\eeq{\end{equation}}
\def\bea{\begin{eqnarray}}
\def\eea{\end{eqnarray}}
\def\mr{\,\mathrm{}}

\preprint{MS-TP-12-12}
\title{Parton densities from LHC vector boson production at
 small and large transverse momenta}
\author{Michael Klasen}
\email[]{michael.klasen@uni-muenster.de}
\author{Matthias Brandt}
\email[]{matthiasbrandt@uni-muenster.de}
\affiliation{Institut f\"ur Theoretische Physik, Westf\"alische
 Wilhelms-Universit\"at M\"unster, Wilhelm-Klemm-Stra\ss{}e 9, D-48149 M\"unster,
 Germany}
\date{\today}
\begin{abstract}
The parton densities of the proton are of fundamental importance not only
for our description of hadronic and nuclear structure, but also for reliable
predictions for new heavy particle searches at colliders. At the large partonic
momentum fractions required for the production of these particles, the
parton distribution functions, in particular the one of the gluon, are
unfortunately still badly constrained. In this paper, we investigate the
possibility to improve on their determination with new data coming from
electroweak vector boson production at large transverse momenta at the LHC
with center-of-mass energies of 7, 8 or 14 TeV.
We demonstrate that this process is dominated by quark-gluon scattering,
that theoretical predictions can be reliably made on the basis of
next-to-leading order perturbation theory and its resummation, and
that these data should thus be used in global fits. We also point out
that the non-perturbative parameters determined from Tevatron run-1
$Z$-boson data at low $p_T$ describe very well the new LHC data at
$\sqrt{s}=7$ TeV.
\end{abstract}
\maketitle


\section{Introduction}

The parton distribution functions (PDFs) of the proton are of fundamental
importance for modern particle physics. Not only do they describe our current
knowledge about the internal structure and symmetries of this basic building
block of matter and represent an important baseline for nuclear structure and
deconfinement studies, but they also enter in the theoretical description of
all hadron collider experiments,
precision determinations of Standard Model parameters and new physics searches,
in particular those at the energy frontier of the Large Hadron Collider
(LHC), as only partonic, but not hadronic cross sections are calculable in
perturbative QCD.

Consequently, a large part of the HERA physics program at DESY \cite{Aaron:2009aa}
and many global analysis efforts \cite{Lai:2010vv,Gao:2013xoa,Martin:2009iq,Ball:2011mu}
have been devoted to improving on this knowledge over
the last decades. Deep-inelastic scattering (DIS), now including a combination of
H1 and ZEUS data from the HERA-1 run \cite{Aaron:2009aa} and complemented at
large values of Bjorken $x$ by older fixed target data
\cite{Benvenuti:1989rh,Benvenuti:1989fm,Arneodo:1996qe,Berge:1989hr,Yang:2000ju,Seligman:1997mc},
still provides the
most important single source of information. It constrains in particular the
(valence) quark densities, as the gluon density enters only at next-to-leading
order (NLO) and can thus only be determined from scaling violations. The
decomposition of the light quark flavours can then be constrained
by neutrino structure function data \cite{Tzanov:2005kr,Onengut:2005kv}
and the strange quark density from dimuon production in neutrino DIS
\cite{Goncharov:2001qe,Bazarko:1994tt}, while the charm structure function $F_2^c$,
and to a lesser extent $F_2^b$ for the bottom quark, are directly accessible at HERA
\cite{Adloff:1996xq,Adloff:2001zj,Aktas:2005iw,Aktas:2004az,Breitweg:1999ad,Chekanov:2003rb,Chekanov:2007ch}.

Traditionally, the gluon density has long been constrained with
prompt photon data \cite{Bonesini:1987bv} through the QCD Compton
process $qg \to q\gamma$. However, photon isolation and fragmentation
uncertainties have proven difficult to overcome
(the situation is only now improving with the advent of the LHC data
\cite{d'Enterria:2012yj}), and long-standing
disagreements in the low-transverse momentum ($p_T$) regime have given rise
to speculations on the importance of an intrinsic $p_T$ of the partons in
the proton \cite{Apanasevich:1997hm}. Recent global analyses 
\cite{Lai:2010vv,Gao:2013xoa,Martin:2009iq,Ball:2011mu} therefore
abstain from the use of prompt-photon data, replacing it with better
understood data (e.g.\ due to new jet algorithms) on inclusive jet
production from DIS at HERA \cite{Chekanov:2002be,Chekanov:2006xr,Aktas:2007aa}
(but so far not from photoproduction \cite{Klasen:1994bj,Klasen:1995ab,Klasen:1996it}
due to the uncertainty on the photon structure function \cite{Albino:2002ck})
and from proton-antiproton collisions at
the Tevatron \cite{Abulencia:2007ez,Aaltonen:2008eq,Abazov:2008ae,Klasen:1996yk,Klasen:1997tj}.
On the other hand, the Drell-Yan like production of electroweak $W$
\cite{Acosta:2005ud,Abazov:2007pm,Abazov:2008qv,Abe:1998rv}
and $Z$ \cite{Abazov:2007jy}
bosons through quark-antiquark fusion to charged leptons and neutrinos
helps to constrain the up and down quark and antiquark densities with
a different weighting than DIS data.\footnote{For a recent study of the
sensitivity of $W^+/W^-$ and $W^{\pm}/Z$ production ratios at the LHC
on the up and down quark PDFs see Ref.\ \cite{Malik:2013kba}.}

In two previous publications that have received considerable attention
\cite{Berger:1998ev,Berger:1999fm}, E.L.\ Berger and one of us (M.K.)
pointed out the possibility to use lepton pairs with relatively
small invariant mass $M$ as a surrogate for prompt photons.
We demonstrated that at intermediate values of $p_T$ of the lepton pair,
its production begins to be dominated by a QCD ``Compton'' process
$qg\to q\gamma^\ast$, with the real photon replaced by a virtual photon
that transforms subsequently into a low-mass lepton pair.
This would allow to use this process for the determination of the gluon
density, also in fixed-target experiments and at large $x$, where it is badly
constrained, while $M$ and $p_T$ are still high enough to allow
for the application of perturbative QCD. This was established by comparing
a calculation with soft-gluon resummation at the next-to-leading logarithmic
(NLL) level to the pure NLO result \cite{Arnold:1990yk}. In a follow-up
publication \cite{Berger:1999es}, we extended this idea to polarized scattering processes,
pointing out the great sensitivity e.g.\ of experiments at RHIC to the largely
unconstrained polarized gluon density.

With the much higher center-of-mass energies of $\sqrt{s}=7$ and 8 TeV that
have become available at the LHC during the last three years and that will
rise to 14 TeV after the current shut-down, it is quite
natural to ask if electroweak boson production can not play a similar role
to virtual photons. The reason is that at such high energies, the $W$- and
$Z$-boson masses of about 80 and 91 GeV will become more and more
negligible, so that quark-gluon scattering should again quickly take
over from quark-antiquark scattering. Therefore, contrary to current
practice, where $W$- and $Z$-boson production are only used in the Drell-Yan
mode at low $p_T$ to constrain the quark flavour decomposition
and in particular the shape of the ratio $d/u$ (down- over up-quark PDFs)
\cite{Gao:2013xoa},
it should soon become possible to also constrain better the gluon density,
in particular at large $x$.

This is the goal of our present work. In Sec.\ 2, we will first review
the current status of parton densities in the proton on the basis of the
uncertainty estimates provided by the three widely used global analyses
CT10 \cite{Lai:2010vv}, MSTW08 \cite{Martin:2009iq} and NNPDF2.1 \cite{Ball:2011mu}.
We will do this at two different scales, i.e.\ at the scale of the electroweak
boson and at a high scale corresponding to large transverse momenta, and emphasizing
both the low-$x$ and high-$x$ regimes. In Sec.\ 3, we will establish the reliability
of our calculation by confronting it to $\sqrt{s}=7$ TeV LHC data on $Z$ and $W$
production with transverse momenta up to 600 and 300 GeV, published recently by the
CMS \cite{Chatrchyan:2011wt} and ATLAS \cite{Aad:2011fp} collaborations, respectively.
We will in particular address the question up to which transverse momenta soft-gluon
radiation must be resummed and at which $p_T$ the NLO perturbative calculation
starts to be reliable. Sec.\ 4 is devoted to an investigation of the different
partonic contributions, i.e.\ at which $p_T$ the quark-gluon process starts to take
over from the Drell-Yan like quark-antiquark process, while in Sec.\ 5 we make
concrete predictions in the perturbative regime for $W$ and $Z$ production up
to large transverse momenta, where the PDFs and in particular the gluon density
can be constrained. For completeness, we also show the sensitivity in the low-$p_T$
regime. Our conclusions and an outlook are given in Sec.\ 6.

\section{Current status of parton density uncertainties}
\label{sec:2}

In this section, we briefly review the current status of parton density uncertainties
in the low- and high-$x$ regimes and their evolution from small to large scales.
As a baseline, we use the best CT10 NLO global fit $f_0$ \cite{Lai:2010vv}, which
we show together with its uncertainty band computed as 
\bea
 \delta^+f&=&\sqrt{\sum_{i=1}^{26}[\max(f_i^{(+)}-f_0,f_i^{(-)}-f_0,0)]^2},\\
 \delta^-f&=&\sqrt{\sum_{i=1}^{26}[\max(f_0-f_i^{(+)},f_0-f_i^{(-)},0)]^2},
\eea
where $f_i^\pm$ are the PDFs for positive and negative variations of the
PDF parameters along the $i$-th eigenvector direction in the 26-dimensional
PDF parameter space. In order to estimate the bias coming, e.g., from
different parameterizations of the $x$-dependence at the starting scale,
we also show the best fits of the MSTW08 \cite{Martin:2009iq} and NNPDF2.1
\cite{Ball:2011mu} global analyses. As it is usually done, we plot in all
cases $x$ times the PDF, i.e.\ the momentum distribution of the partons
in the proton.

\begin{figure}[!h]
 \centering
 \epsfig{file=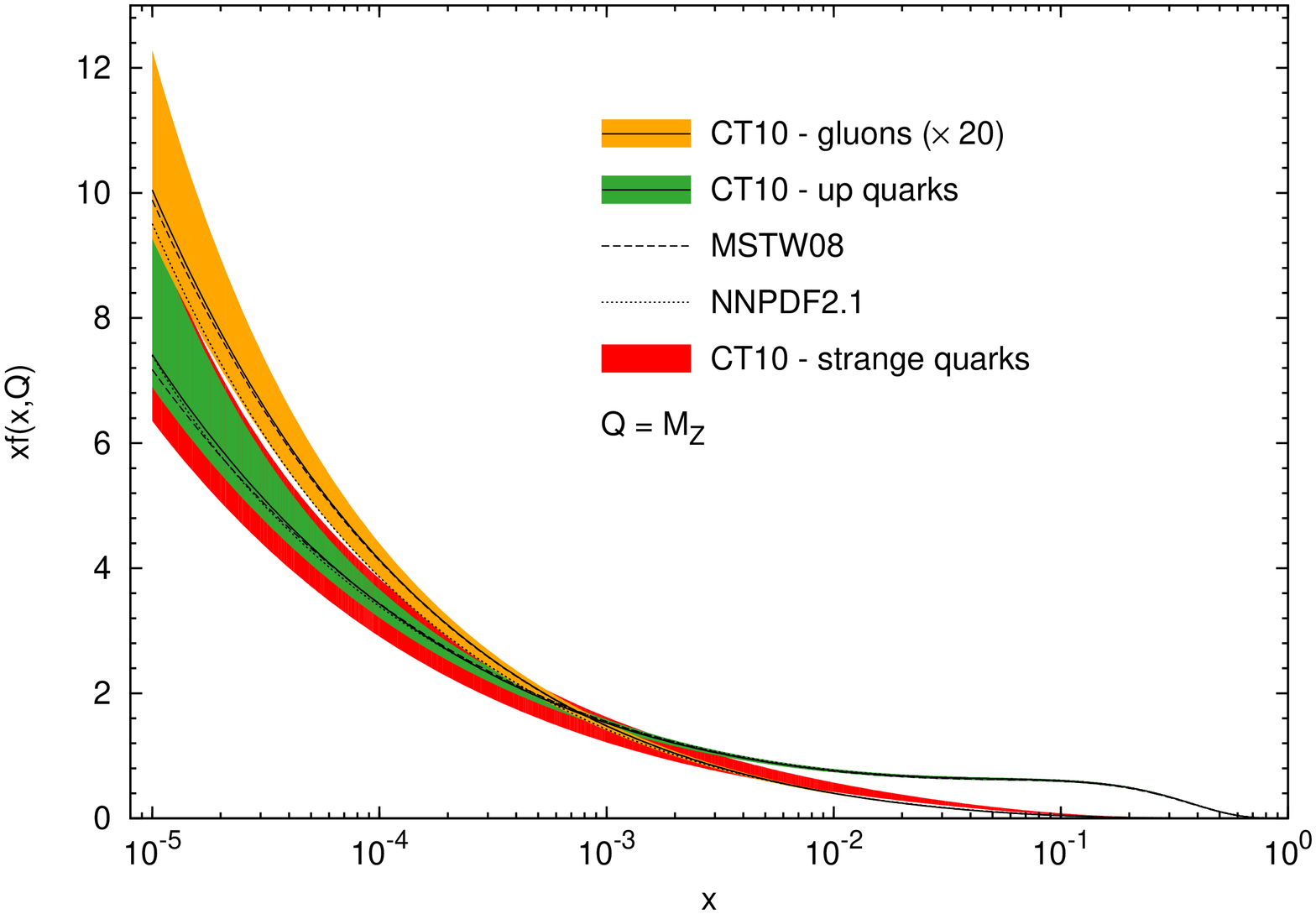,width=0.63\columnwidth,angle=-90}
 \epsfig{file=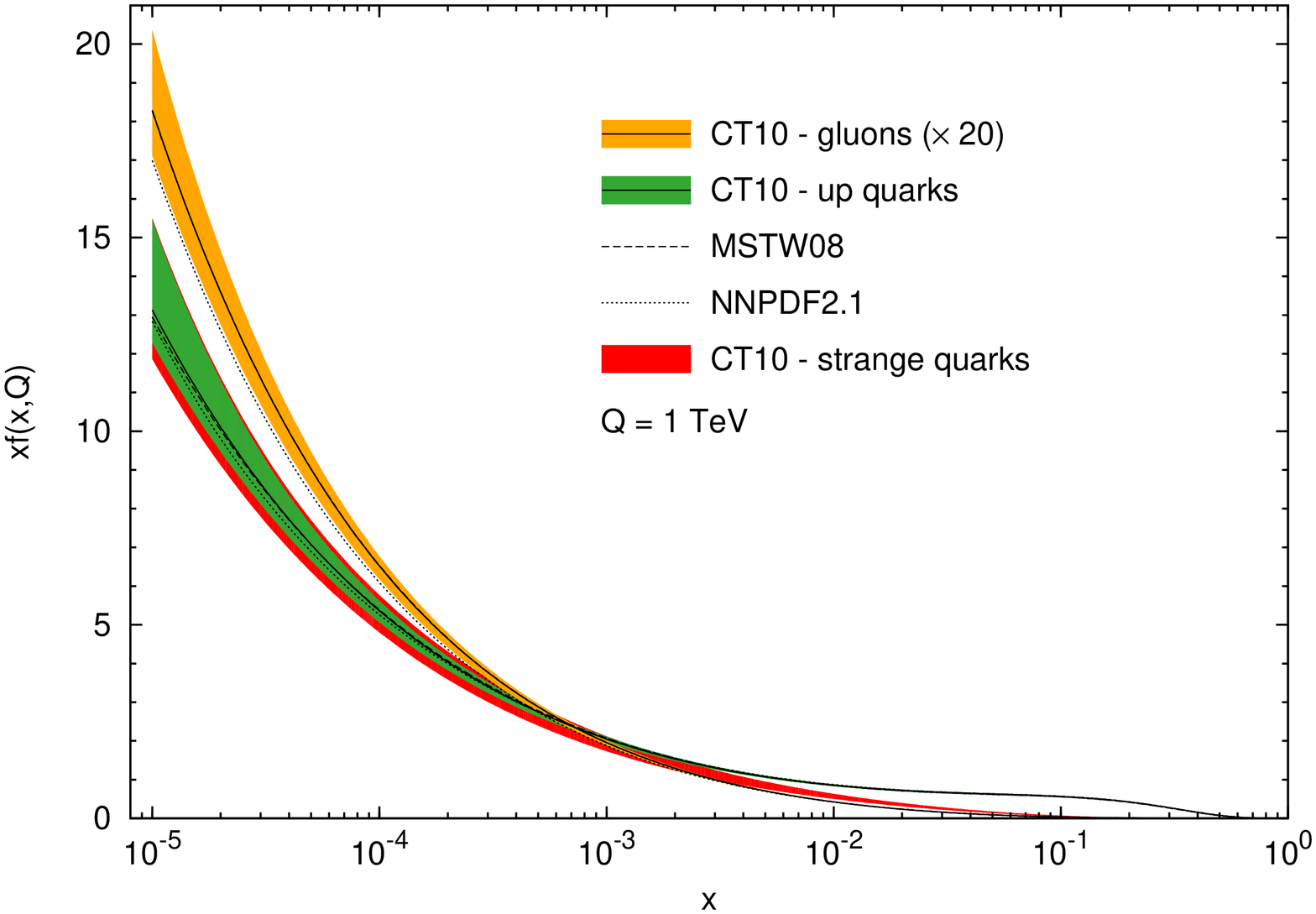,width=0.63\columnwidth,angle=-90}
 \caption{\label{fig:1}Parton distribution functions (PDFs) on a logarithmic $x$-scale,
 emphasizing the low-$x$ region, at the factorization scale
 $Q = M_Z$ (top) and 1 TeV (bottom) from different collaborations.
 PDF uncertainties are only shown for the CT10 group. PDFs are printed for gluons, up and
 strange quarks.}
\end{figure}

First, we show in Fig.\ \ref{fig:1} the gluon PDF (reduced by a factor of 20),
%
%
the up-quark and the strange-quark PDFs at the factorization scale $Q=M_Z$ (top),
adequate for electroweak boson production at low transverse momenta, and
at a higher scale of $Q=1$ TeV (bottom), relevant for high-$p_T$ vector boson
production. In this figure, a logarithmic $x$-axis has been chosen, which
emphasizes the low-$x$ regime. In this region, the sea-quark dominated
up- and strange-quark PDFs largely overlap, so that for better visibility no
central values, but only the uncertainty band is shown for the latter. As 
is well known, the uncertainty highly increases below values of $x=10^{-3}$,
where only little information is available from pre-LHC experiments. The
evolution from low (top) to high (bottom) scales resums multiple parton
splitting, increasing the densities of gluons and sea quarks at small $x$
and at the same time reducing the (mostly valence-quark) PDFs at large $x$.
Since the $Q^2$-dependence is perturbatively calculable, the PDFs at
high $Q$ become less dependent on the non-perturbative input at
the starting scale $Q_0$, so that their uncertainty, due to fit of the
unknown $x$-dependence at $Q_o$ to experimental data, is reduced.

\begin{figure}[!h]
 \centering
  \epsfig{file=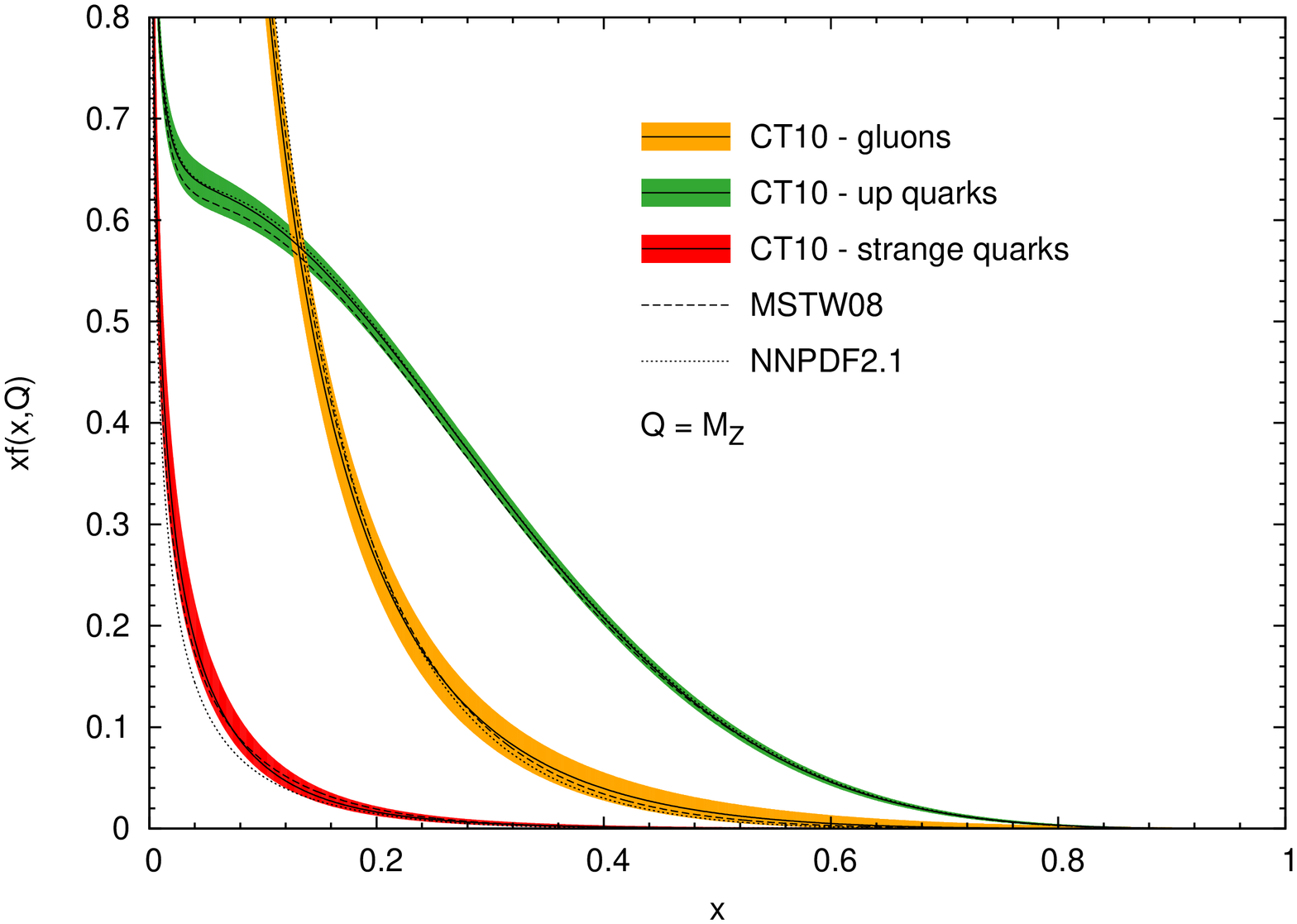,width=0.63\columnwidth,angle=-90}
  \epsfig{file=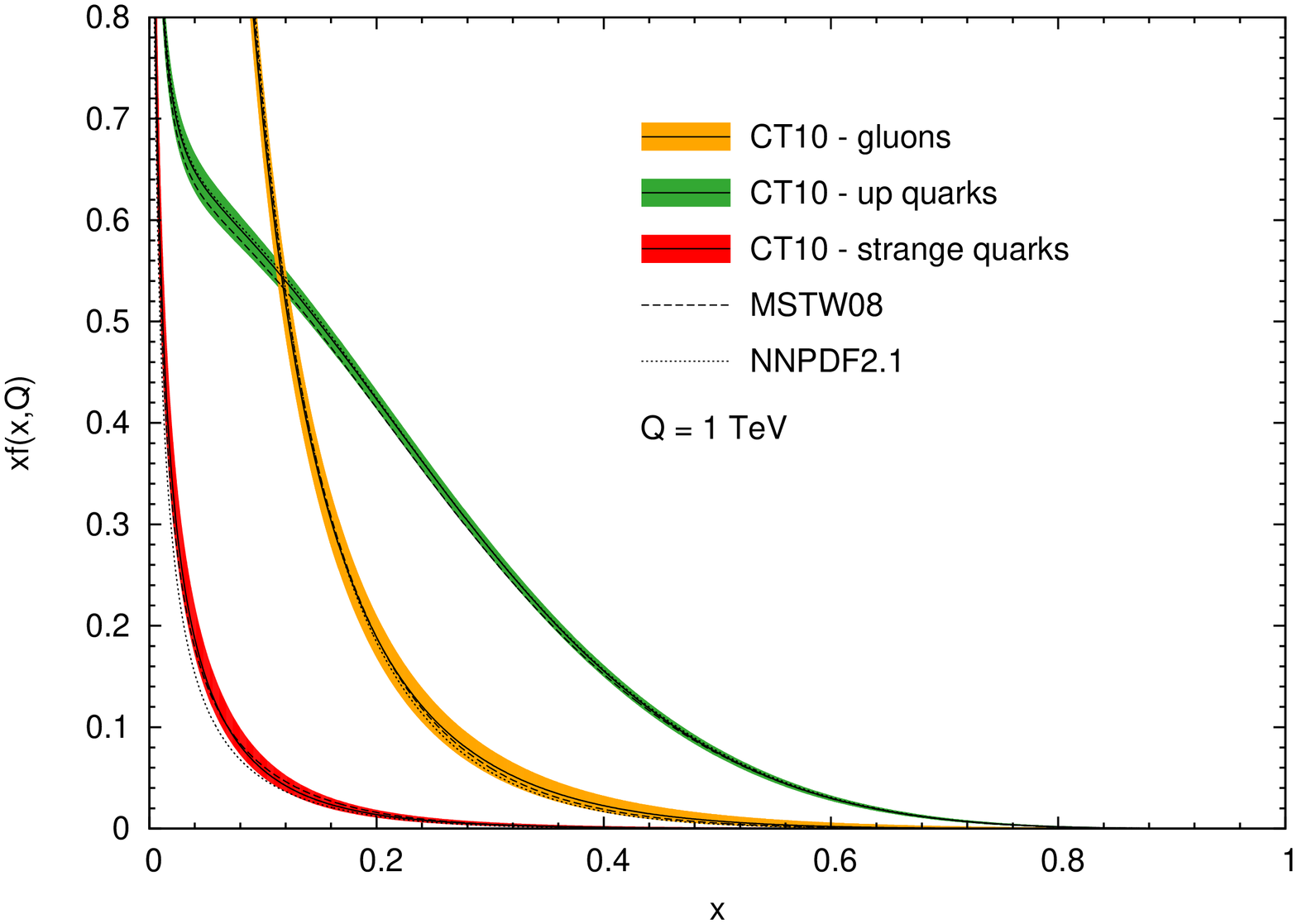,width=0.63\columnwidth,angle=-90}
 \caption{\label{fig:2}Same as in Fig.\ \ref{fig:1}, but on a linear $x$-scale,
 emphasizing the high-$x$ region.}
\end{figure}

The shift of the up-quark, but also the gluon (and induced strange-quark)
PDFs from larger to smaller $x$-values with the evolution from $Q=M_Z$ to
$Q=1$ TeV is more clearly visible in Fig.\ \ref{fig:2} with its linear $x$-axis.
%
%
Here, up quarks (which have a valence-quark contribution) and strange quarks
(which do not) are well separated, so that we now also show the central values
for the latter. It is obvious that the CT10 uncertainty band, induced by the
experimental data, does not cover in all cases the central values of MSTW08
and NNPDF2.1, demonstrating the important influence of the theoretical bias
on the $x$-parametrization at the starting scale $Q_0$.

If we focus now on the gluon PDF at large $x$, we can see that it is much
less constrained than the up-quark PDF, directly probed at HERA and in
other DIS experiments. At the lower scale $Q=M_Z$, the gluon uncertainty
parametrized by CT10 amounts at $x=0.3$ and $x=0.4$ to considerable
+21/-18 \% and +40/-30 \%, respectively. The evolution to $Q=1$ TeV does
not significantly change this uncertainty, amounting there to +22/-17 \% and
+43/-28 \%, respectively. Contrary to the strange-quark PDFs, the central
MSTW08 and NNPDF2.1 gluon PDF fits are covered by the CT10 uncertainty bands.

In the three global analyses discussed above, LHC data have not yet been
included. However, the new CT10 next-to-next-to-leading order (NNLO)
PDFs \cite{Gao:2013xoa} have been
compared with the total $W$ and $Z$ cross sections at 7 and 8 TeV with the
result that they agree within the still substantial errors. Also the
$W$ and $Z$ rapidity distributions as well as the $W$ charge asymmetry at
7 TeV have been found to agree within errors. These data are obtained in
a Drell-Yan situation at small $p_T$ and, when included in future global
PDF analyses, will mostly influence the quark PDFs at small $x$, in
particular the in this region badly constrained valence quarks. This requires,
however, the theoretical predictions for the partonic scattering cross
sections including soft-gluon resummation to be under control \cite{Gao:2013xoa}.

\section{NLO cross sections with resummation for the LHC}
\label{sec:3}

As our previous calculations for low-mass lepton-pair production in hadron
collisions \cite{Berger:1998ev,Berger:1999fm}, our theoretical predictions for massive
electroweak gauge-boson production at the LHC are based on a full NLO
calculation matched to soft-gluon resummation at the NLL level
\cite{Arnold:1990yk}. This allows us to establish in this section the
regions in $p_T$ where the perturbative results are reliable and where
they have to be supplemented by a resummation procedure to all orders
in the strong coupling constant $\alpha_s$. This $p_T$-resummation
is based on the Collins-Soper-Sterman formalism \cite{Collins:1984kg}
and is evaluated in impact parameter ($b$) space. The differential
cross section for the production of a vector boson $V$ from two hadrons $h_1$
and $h_2$ is thus written as
\beq
 {\d\sigma(h_1h_2\to VX)\over\d Q^2\d p_T^2\d y}={1\over(2\pi)^2}\delta(Q^2-M_V^2)
 \int\d^2b\,e^{i\vec{p}_T\vec{b}}\,\tilde{W}_{j\bar{k}}(b,Q,x_1,x_2)+Y(p_T,Q,x_1,x_2),
\eeq
where $Q$ and $y$ are the invariant mass and rapidity of the vector boson
and $x_{1,2}=e^{\pm y}Q/\sqrt{s}$ are the momentum fractions of the interacting partons
$j$ and $\bar{k}$. The regular piece, denoted $Y(p_T,Q,x_1,x_2)$, is obtained by subtracting
the terms, which are singular in $p_T$, from the exact fixed-order result.
The form factor $\tilde{W}_{j\bar{k}}(b,Q,x_1,x_2)$ can be factorized into a 
perturbative piece $\tilde{W}_{j\bar{k}}^{\rm pert}(b_{\ast})$ and a non-perturbative
function $\tilde{W}_{j\bar{k}}^{\rm NP}(b)$,
\bea
 \tilde{W}_{j\bar{k}}(b,Q,x_1,x_2)&=& \tilde{W}_{j\bar{k}}^{\rm pert}(b_{\ast},Q,x_1,x_2)
 \tilde{W}_{j\bar{k}}^{\rm NP}(b,Q,x_1,x_2)
 \label{eq:3.2}
\eea
when introducing a variable
\bea
 b_{\ast}&=&{b\over \sqrt{1+(b/b_{\max})^2}},
\eea
which, together with $b_{\max}=0.5$ GeV$^{-1}$, ensures the perturbativity of
$\tilde{W}_{j\bar{k}}^{\rm pert}$. 
A renormalization group analysis \cite{Collins:1984kg} exhibits a logarithmic dependence of the
non-perturbative function $\tilde{W}_{j\bar{k}}^{\rm NP}(b,Q,Q_0,x_1,x_2)$ on the
starting scale $Q_0$ of the PDFs, which turns, however,
out to be negligible in practice. Its $b$- and $x$-dependence must be fitted
to experimental data. We use a Gaussian parametrization by
Brock, Landry, Nadolsky and Yuan (BLNY)
\bea
 \tilde{W}^{\rm NP}_{j\bar{k}}(b,Q,Q_0,x_1,x_2)&=&
 \exp\le -g_1-g_2\ln\lr {Q\over2Q_0}\rr-g_1g_3\ln(100\,x_1x_2)\re b^2
\eea
with the three parameters $g_1=0.21$, $g_2=0.68$ and $g_3=-0.60$ and
evolved from the starting scale $Q_0=1.6$ GeV, which has been shown to fit
the Tevatron run-1 data on $Z$-boson production very well \cite{Landry:2002ix}.

\begin{figure}[!h]
 \centering
  \epsfig{file=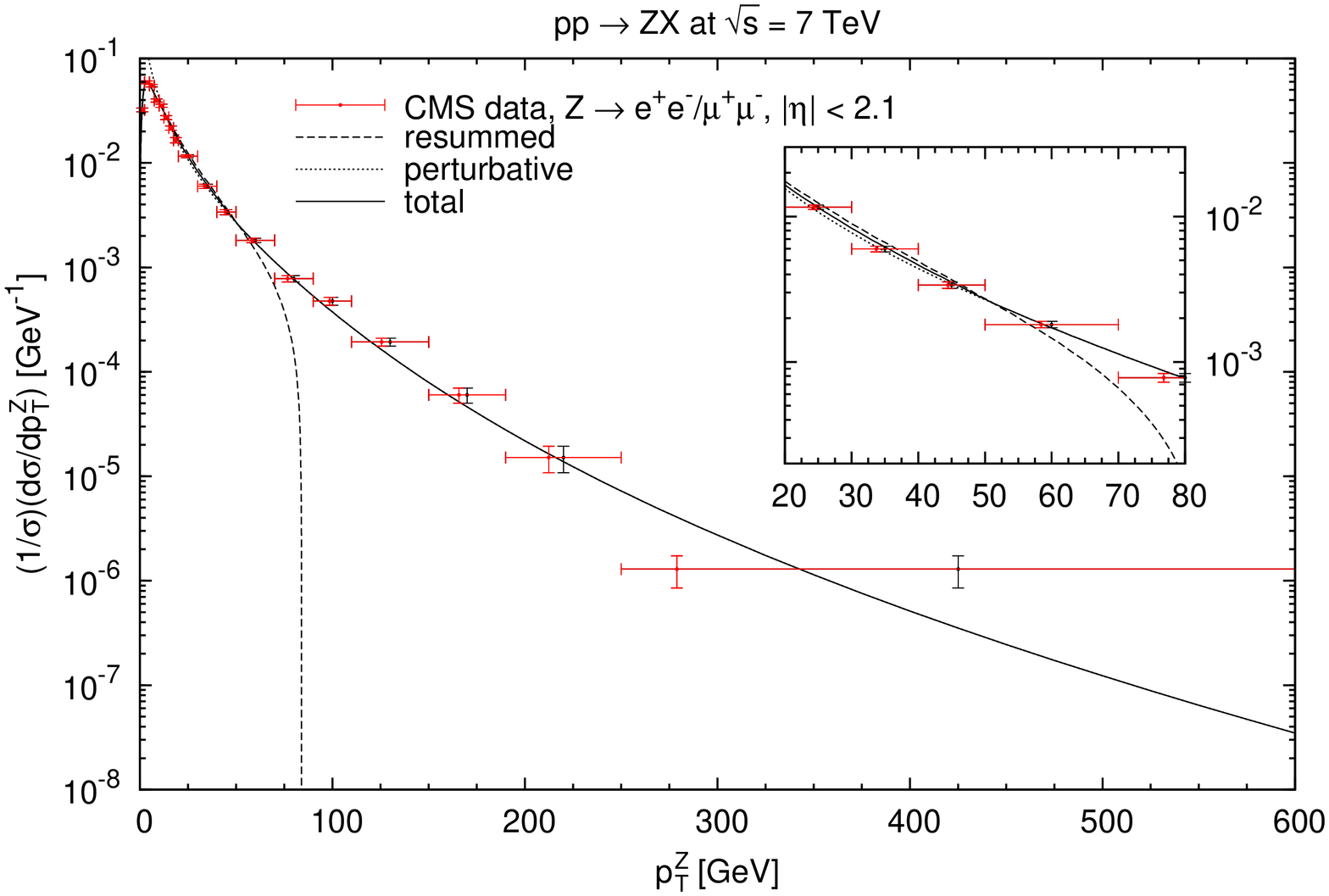,width=0.63\columnwidth,angle=-90}
  \epsfig{file=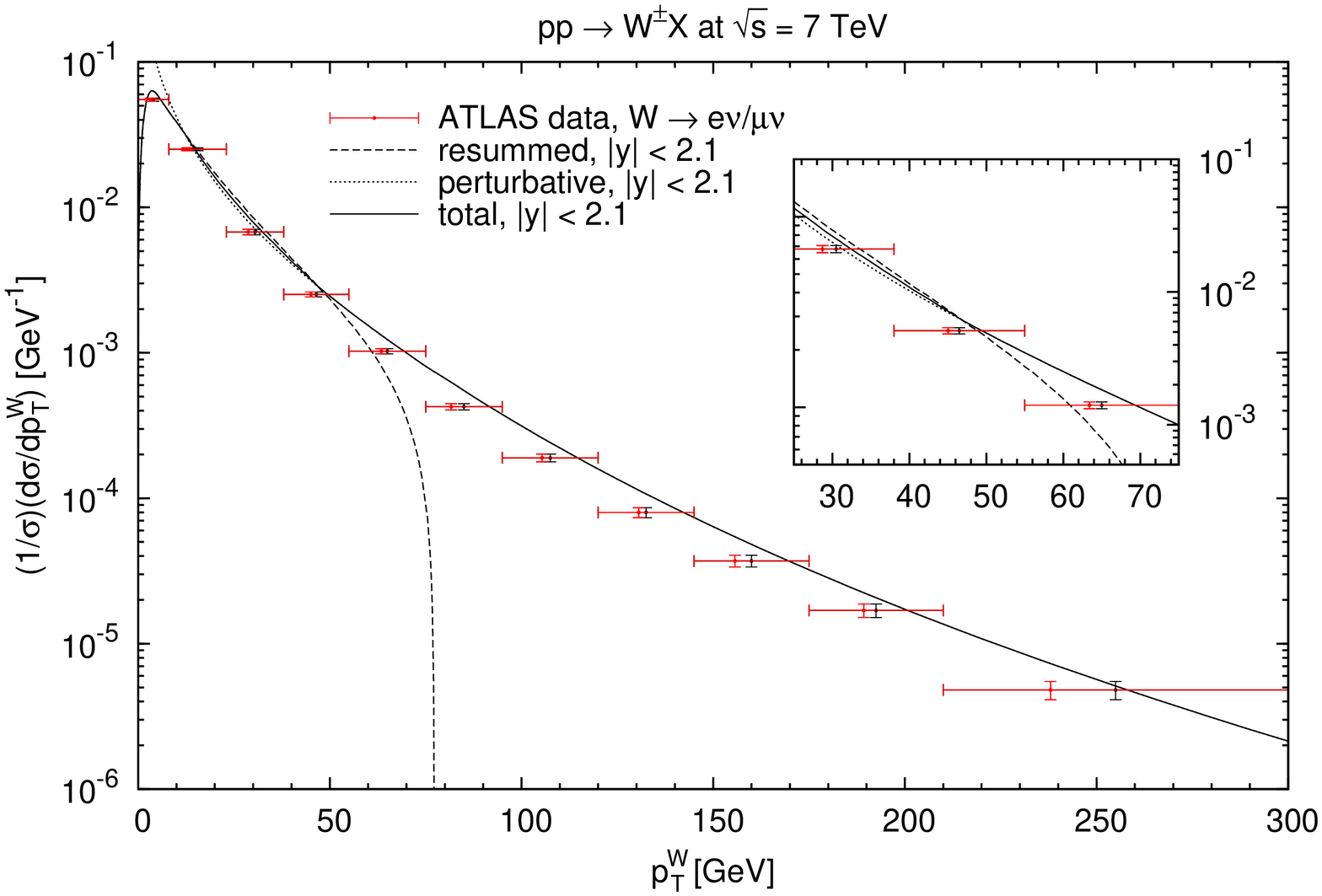,width=0.63\columnwidth,angle=-90}
 \caption{\label{fig:3}Transverse-momentum spectra of $Z$ (top) and $W$ (bottom)
 bosons at the LHC with $\sqrt{s}=7$ TeV, normalized to the total cross section.
 CMS (top) and ATLAS (bottom) data are compared with
 our theoretical calculation at NLL+NLO in the rapidity range $|y|<2.1$ using CT10 PDFs.
 The data points are positioned at the theoretical center of gravity of the bins (red)
 and at the center of the bins (black).}
\end{figure}

Our theoretical predictions computed in this way are compared in the upper part
of Fig.\ \ref{fig:3} to CMS data on $Z$-boson production at $\sqrt{s}=7$ TeV
%
%
\cite{Chatrchyan:2011wt}. The data were presented by the CMS collaboration normalized
to the total cross section and with combined statistical and systematic uncertainties,
individually for decay electrons and muons with $|\eta|<2.1$ and as a combination.
They extend to $p_T$ values of 600 GeV for the $Z$-boson. We only show the combination
in Fig.\ \ref{fig:3}, where we have multiplied our $Z$-boson cross section, computed
with our baseline PDF set CT10 \cite{Lai:2010vv} and integrated over $|y|<2.1$, with
the relevant branching fractions \cite{Beringer:1900zz}. As one can observe, the
region, where resummation is needed to describe the data, extends to values of
$p_T\simeq 75$ GeV. Below this point, the perturbative calculation (dotted) diverges
logarithmically due to multiple soft-gluon radiation and must be resummed (dashed),
while above the regular, non-logarithmic terms due to hard, non-collinear radiation
can no longer be neglected as it is done in the resummation calculation. The failure
of the soft-gluon approximation in the transition region is exhibited by the fact
that the resummation prediction becomes negative there and must be matched to the
perturbative result by re-expanding it, subtracting from it the divergent terms,
and then adding the perturbative result to obtain a prediction valid in all
regions (full curve). The comparison of our theoretical predictions with the
experimental data is excellent over the full region in $p_T$. Similarly good
agreement has been found by the CMS collaboration with predictions based on the
POWHEG NLO Monte Carlo generator with parton showers, which effectively also
resum the leading logarithms at small values of $p_T$ \cite{Frixione:2007vw,Alioli:2008gx}.

In the lower part of Fig.\ \ref{fig:3}, we compare our theoretical predictions
to ATLAS data on $W$-boson production at $\sqrt{s}=7$ TeV \cite{Aad:2011fp}.
Similarly to CMS,
the ATLAS collaboration present their results normalized to the total cross
section and for a combination of weak bosons
decaying to electrons and muons, measured now with $|\eta|<2.4$. The decay
neutrinos of course escape detection. The $p_T$-spectrum of the $W$ boson is
then obtained through a two-step unfolding procedure up to values of 300 GeV.
The theoretical behaviour
is very similar to the one for $Z$-boson production, except that the transition
from the region dominated by large logarithms to the perturbative regime occurs
at slightly smaller values of $p_T\simeq65$ GeV. This can be attributed to the
other hard scale in the process, the mass of the $W$ boson, which is with 80.385 GeV
somewhat smaller than the $Z$-boson mass of 91.188 GeV \cite{Beringer:1900zz}. 
If we assume the rapidity region of the $W$-boson, not given explicitly by the
ATLAS experiment, to be $|y|<2.1$, we obtain again very good agreement with
the experimental data, although the normalization to the total cross section
renders the prediction almost insensitive to the exact value of this cut.
The experimentalists obtain similarly good agreement when comparing to the
resummation program RESBOS \cite{Ladinsky:1993zn,Balazs:1997xd,Landry:2002ix}.

An important question is how the resummation and perturbative regions change
when moving from current LHC experiments with $\sqrt{s}=7$ TeV to those in
the future, which will be conducted with collision energies up to 14 TeV.
We have investigated this question with the result (not shown explicitly)
that for $Z$ bosons the pure resummation result then starts to deviate
strongly from the total prediction at values of $p_T\simeq90$ GeV,
whereas for $W$ bosons this point is reached at $p_T\simeq 80$ GeV.
On the other hand, our calculations indicate that the reach in $p_T$,
which was 600 (300) GeV for $Z$ ($W$) bosons produced at $\sqrt{s}=7$
TeV with an integrated luminosity of 35.9 (31) pb$^{-1}$, should increase
to the multi-TeV range, i.e.\ at least 2 TeV, at $\sqrt{s}=14$ TeV and an
integrated luminosity of 100 fb$^{-1}$. At the end of 2012, already
more than 23 fb$^{-1}$ had been recorded by ATLAS and CMS each with
$\sqrt{s}=8$ TeV. It would thus already be interesting to analyse these
data for electroweak vector boson production with high $p_T$.


\section{Decomposition of partonic processes}

The next question that arises is then at which values of $p_T$
electroweak boson production starts to be dominated by the QCD ``Compton''
processes $qg\to Zq$ and $qg\to Wq$, just as low-mass lepton pairs
(with invariant mass below $M_W$) were dominated by virtual photon
radiation through the process $qg\to\gamma^{\ast}q$.\\

\begin{figure}[!h]
 \centering
  \epsfig{file=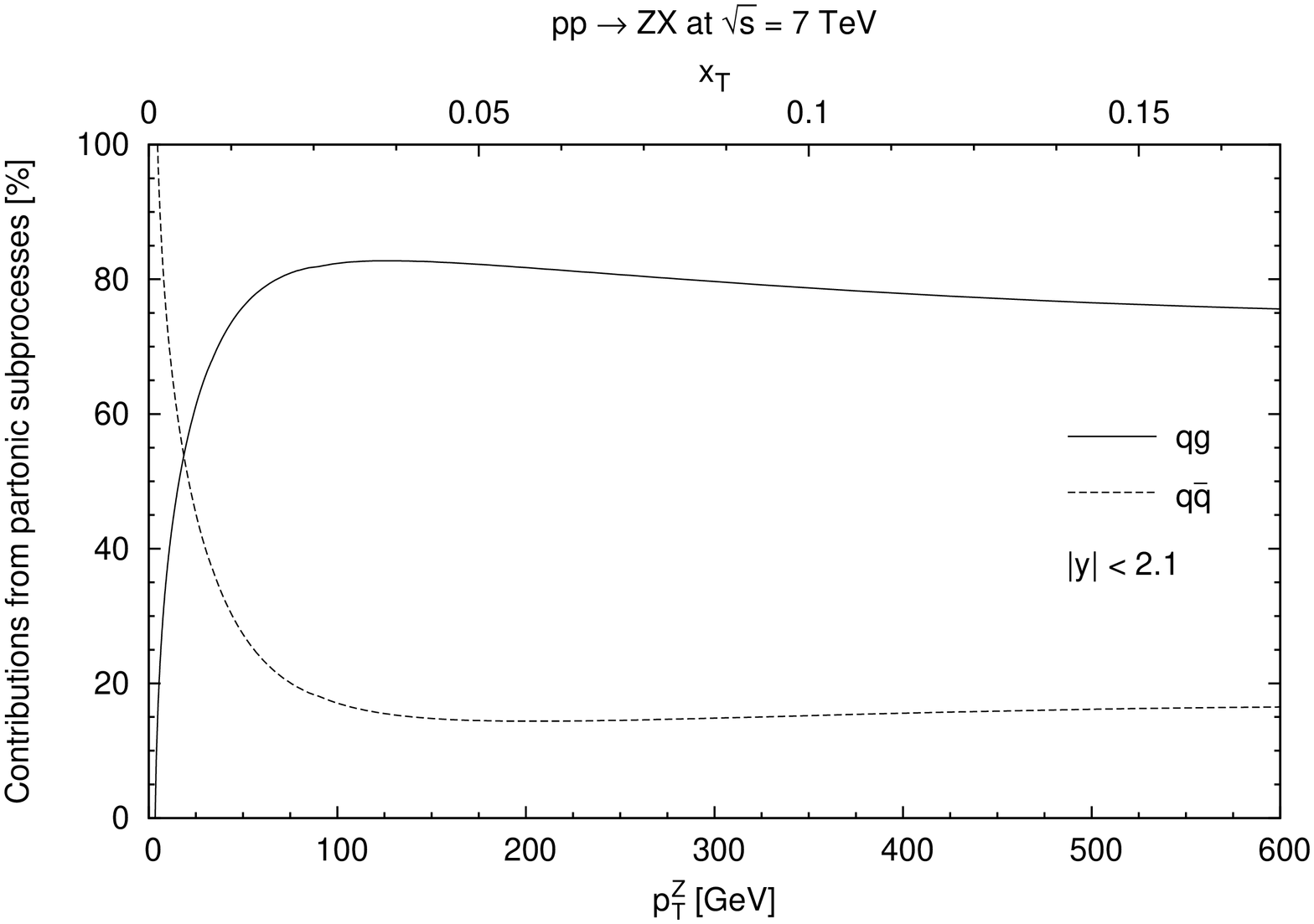,width=0.64\columnwidth,angle=-90}
  \epsfig{file=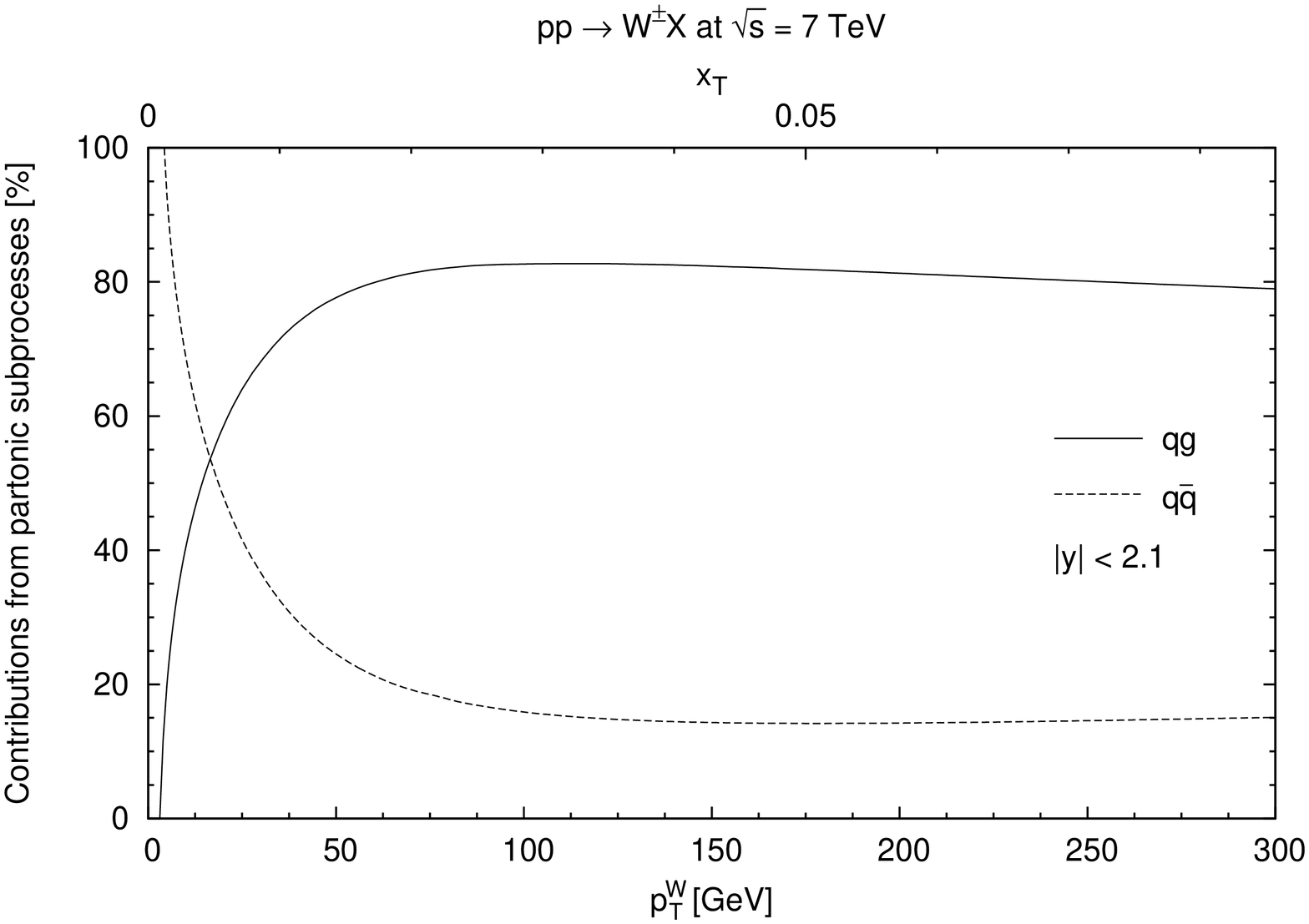,width=0.64\columnwidth,angle=-90}
 \caption{\label{fig:4}Relative contributions at NLO of the quark-antiquark (dashed) and
 QCD ``Compton'' (full) subprocesses to the production of $Z$ (top) and $W$ bosons (bottom)
 at the LHC with $\sqrt{s}=7$ TeV. Subdominant partonic subprocesses which enter only at
 NLO or higher order are not shown.}
\end{figure}

The importance of the quark-gluon scattering process is clearly visible
in Fig.\ \ref{fig:4}. In fact, at a proton-proton collision energy of
%
%
$\sqrt{s}=7$ TeV it is sufficient for the transverse momentum of the produced
$Z$ (top) or $W$ (bottom) boson to exceed 20 or 15 GeV. The quark-gluon process
then remains at a level of 75-80\% almost up to the kinematic limit, more
precisely up to 3 TeV, before quark-antiquark fusion takes over again.
Note that in this figure we show only the subprocesses that exist already
at leading order. At NLO and beyond, other processes like
$gg\to Vq\bar{q}$ and $qq\to Vqq$ with $V=Z,W$ enter as well, but they remain at the
level of a few percent. This statement depends of course on the factorization
scheme and scale, which we choose to be the $\overline{\rm MS}$ scheme and
$\mu_F=\sqrt{M_V^2+p_T^2}$, identical to the renormalization scale $\mu_R$.
This choice should in principle provide for an optimal stability of the
perturbative calculation in the low- and high-$p_T$ regions.\\

\begin{figure}[!h]
 \centering
  \epsfig{file=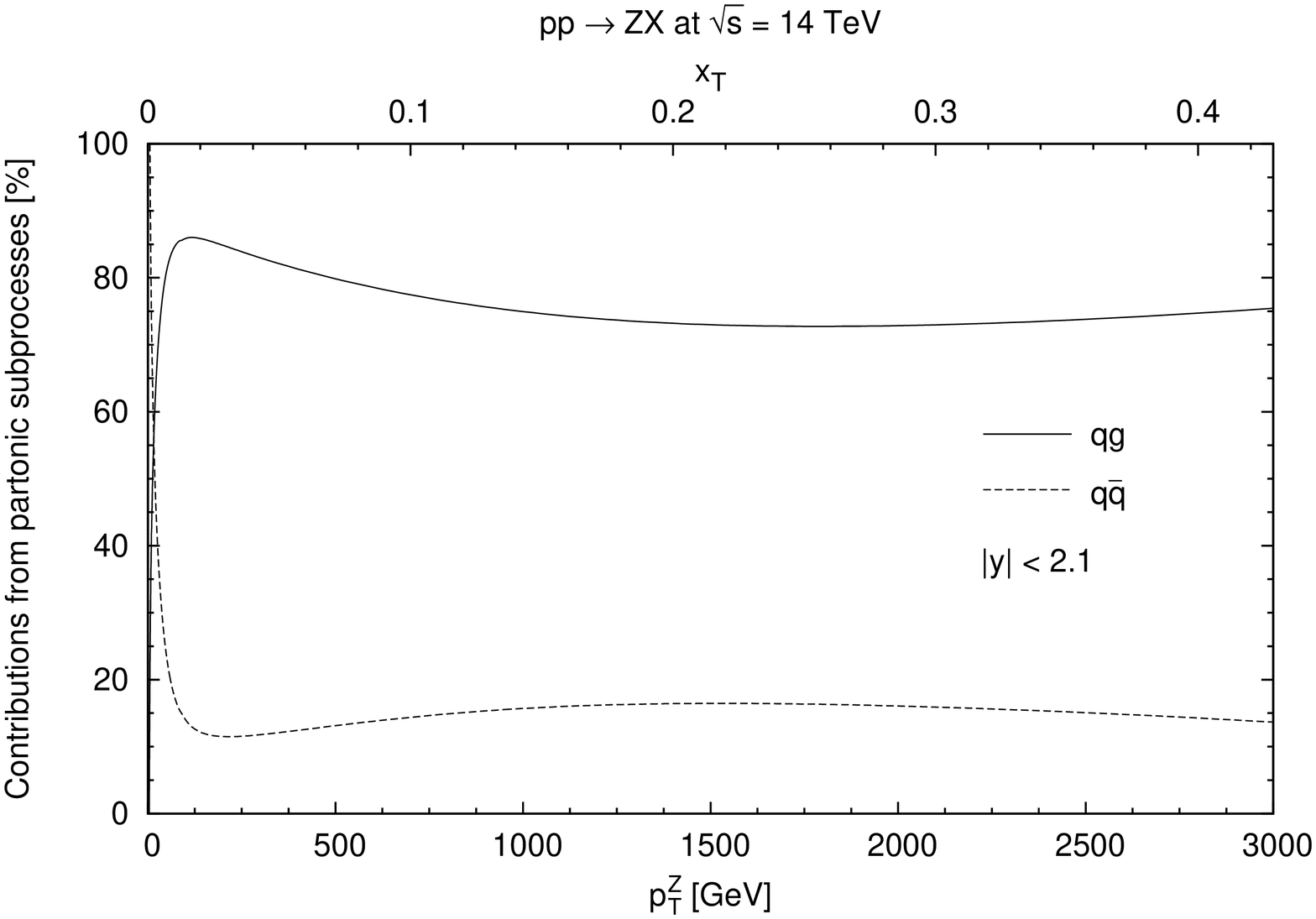,width=0.64\columnwidth,angle=-90}
  \epsfig{file=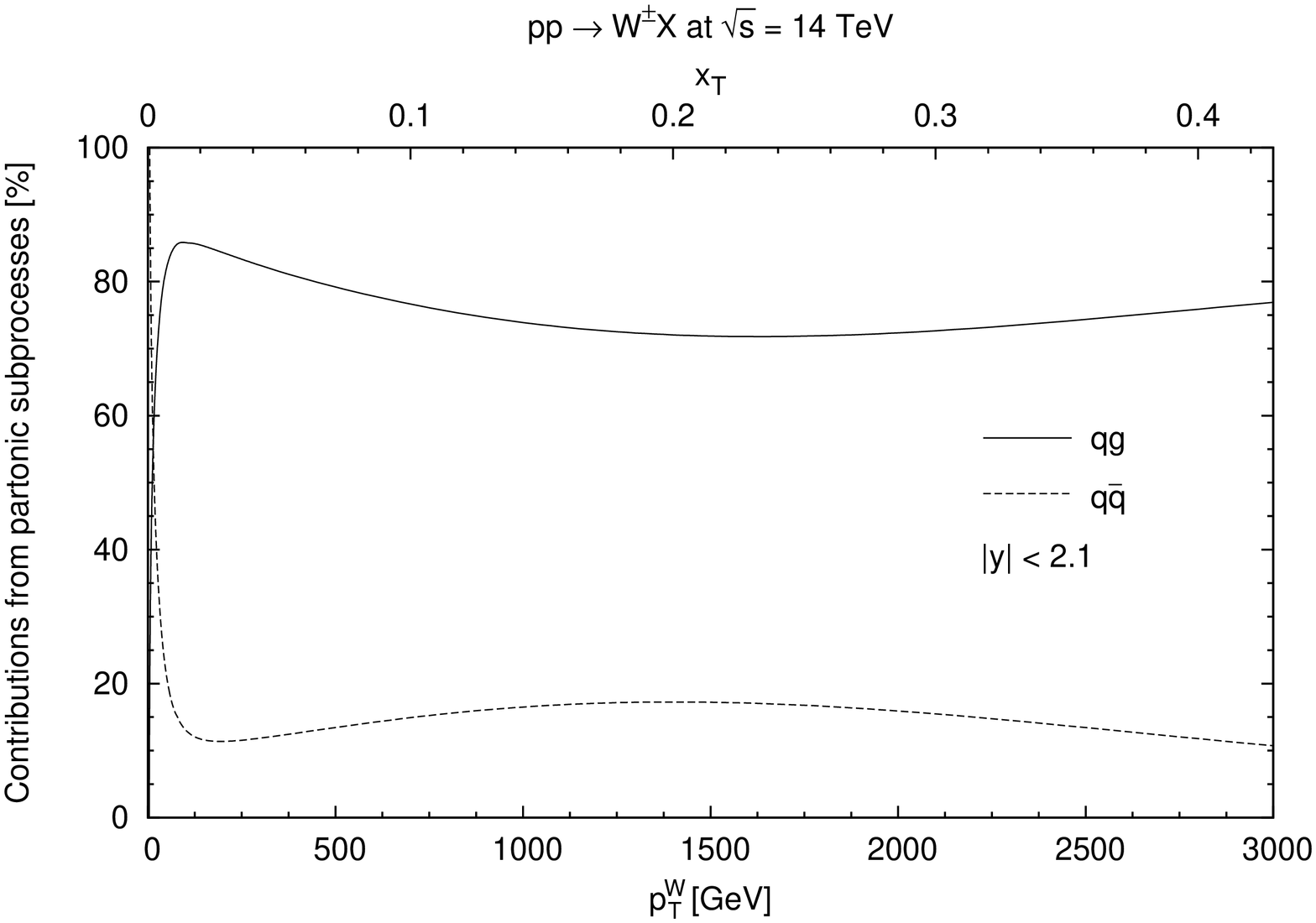,width=0.64\columnwidth,angle=-90}
 \caption{\label{fig:5}Same as Fig.\ \ref{fig:4} for $\sqrt{s}=14$ TeV.}
\end{figure}

The dominance of the $qg$ subprocess persists at higher collision energies
of $\sqrt{s}=14$ TeV, as can be seen from Fig.\ \ref{fig:5}.
%
%
The peak contribution is even a bit larger and reaches more than 85\% at
$p_T\simeq 100$ GeV. A local minimum of about 70\% exists at intermediate values
of $p_T\simeq 1.5$ TeV. In Figs.\ \ref{fig:4} and \ref{fig:5} we have in
addition introduced a second, upper $x$-axis. It shows an estimator for
the values of Bjorken-$x$, more precisely $x_T=2p_T/\sqrt{s}$, at which
the parton distributions in the colliding protons are probed. It is clear
that at the very large values of $p_T$ accessible with high LHC luminosities
and energies of 14 (and also 8) TeV, it should be possible to
probe and constrain the gluon density where it is not well known.

\section{Parton density sensitivity of LHC vector boson production}

Having established the reliability of our calculations in
the resummation and perturbative regimes as well as the dominance of
the quark-antiquark and quark-gluon subprocesses at small and
intermediate to large transverse momenta, we can
now confront the current status of uncertainties on the quark and
gluon PDFs in the proton with the prospects for improving on their
determination with electroweak boson production at the LHC.

\begin{figure}[!h]
 \centering
  \epsfig{file=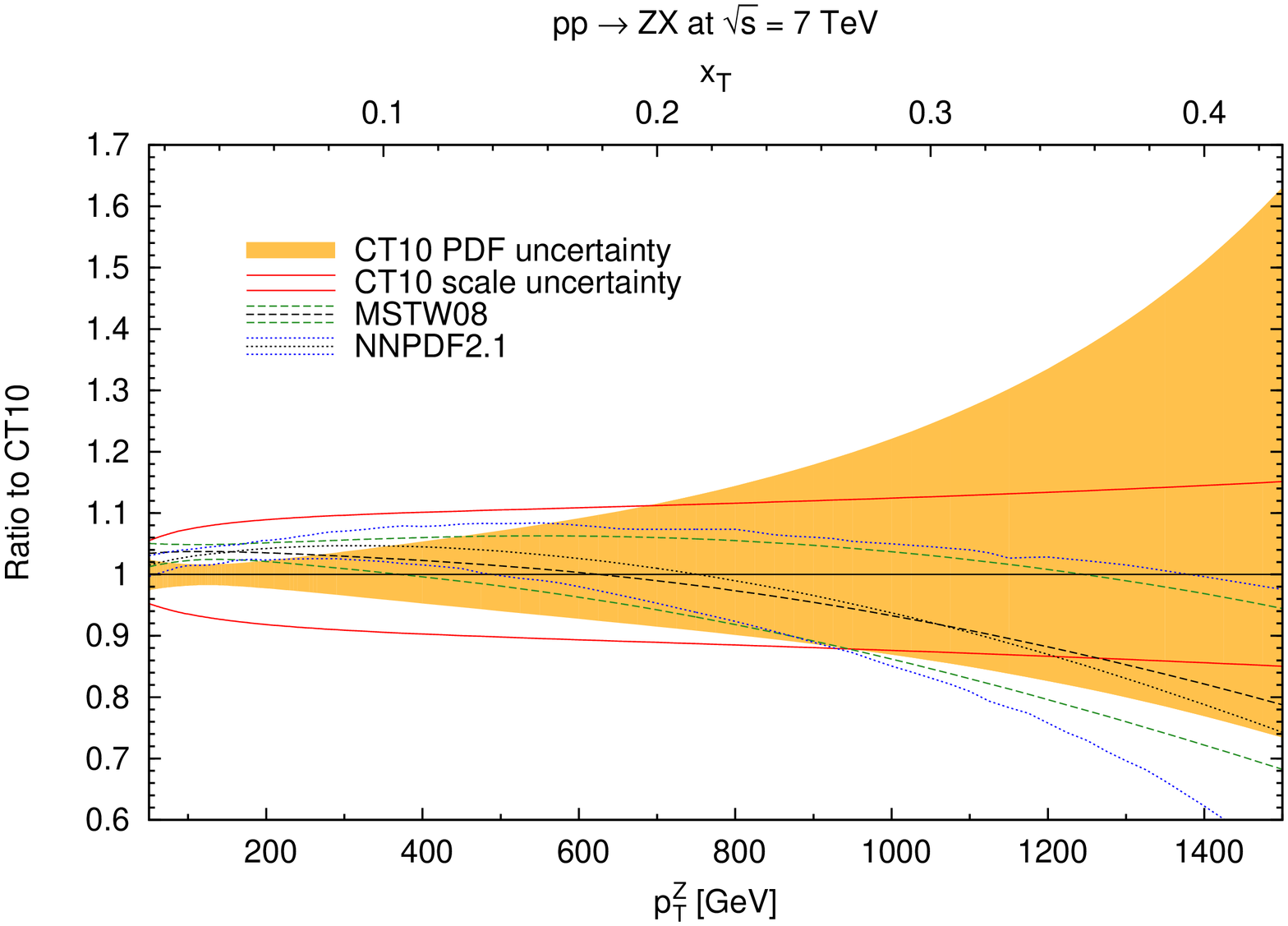,width=0.66\columnwidth,angle=-90}
  \epsfig{file=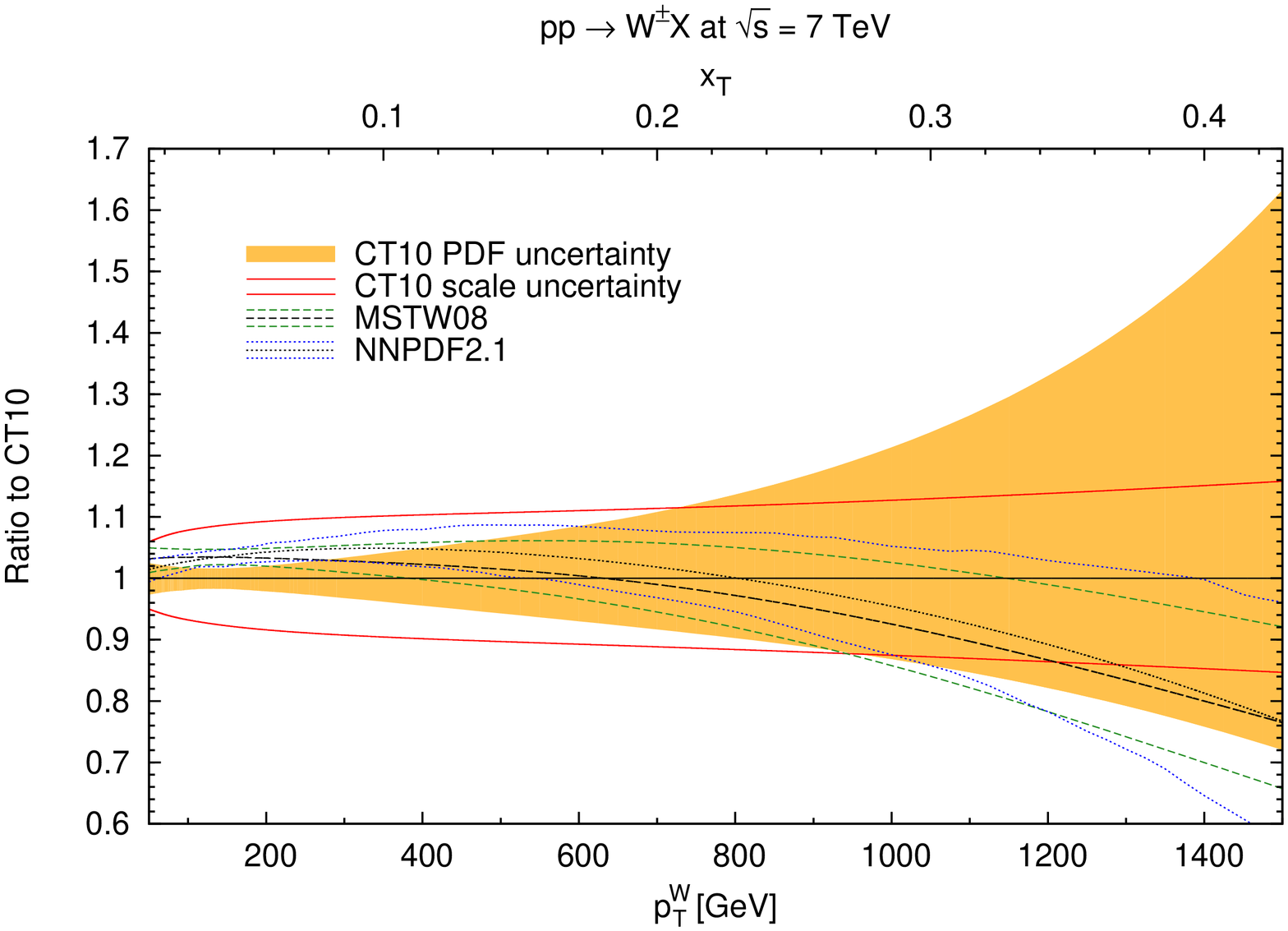,width=0.66\columnwidth,angle=-90}
 \caption{\label{fig:6}Ratios of $Z$- (top) and $W$-boson (bottom) $p_T$-distributions at
 $\sqrt{s}=7$ TeV
 computed with different scales and PDFs to the baseline predictions with CT10 and central
 scale.}
\end{figure}

\begin{figure}[!h]
 \centering
  \epsfig{file=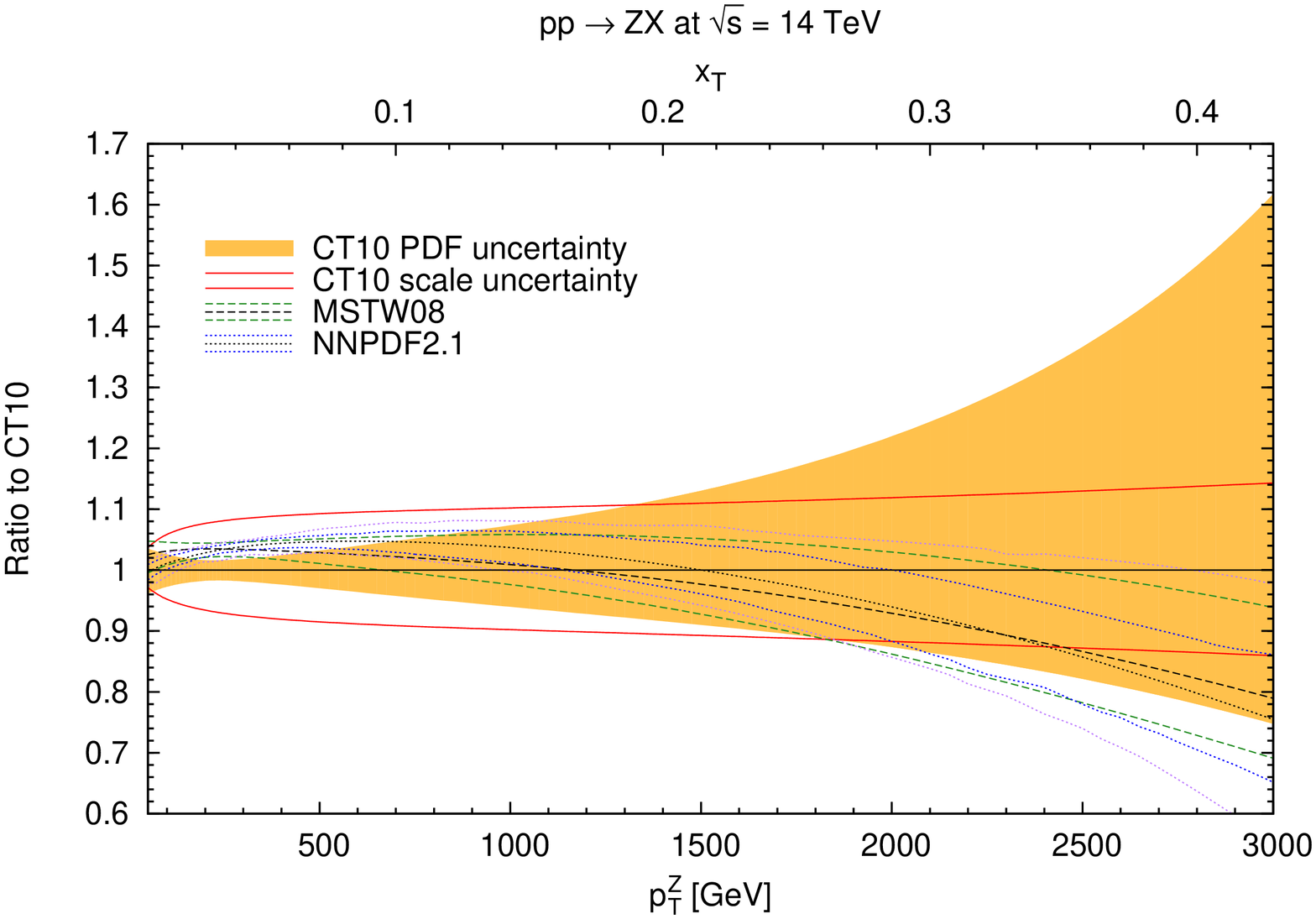,width=0.65\columnwidth,angle=-90}
 \caption{\label{fig:7}Same as Fig.\ \ref{fig:6} for $Z$ bosons at $\sqrt{s}=14$ TeV.}
\end{figure}

To this end, we compute in Fig.\ \ref{fig:6} ratios of transverse-momentum
%
%
spectra for $Z$ (top) and $W$ (bottom) bosons using various PDFs to our
baseline prediction with CT10 PDFs. While this figure shows results for
the LHC with $\sqrt{s}=7$ TeV, Fig.\ \ref{fig:7} shows these ratios for
%
%
$Z$-boson production at a center-of-mass energy of $\sqrt{s}=14$ TeV.
If one accounts for a rescaling of transverse momenta by a factor of
two, these figures are very similar. Results for $\sqrt{s}=8$ TeV are
therefore not shown, as they lie naturally in between, close to the
results of $\sqrt{s}=7$ TeV. With a total luminosity of more than
23 fb$^{-1}$ collected by ATLAS and CMS each at $\sqrt{s}=8$ TeV, a
range in $p_T$ up to 1 TeV or more can be expected. This corresponds
to values of Bjorken-$x$ or, more precisely $x_T$ (upper $x$-axis),
of about 0.3 or more.

In our discussion in Sec.\ \ref{sec:2}, we had observed that the gluon
uncertainty at $x=0.3$ and $x=0.4$ parametrized by CT10 is
quite large and reaches at the scale $Q=1$ TeV values of +22/-17 \% and +43/-28 \%,
respectively. Since the QCD ``Compton'' process contributes here
more than 75\% to the total cross section, this uncertainty is
directly reflected in Figs.\ \ref{fig:6} and \ref{fig:7} through
the yellow CT10 uncertainty bands. The quark PDFs are dominated
in this region by the valence contribution and add only little
to the total uncertainty. The alternative PDF determinations by
MSTW08 (dashed) and NNPDF2.1 (dotted) follow in this region the
lower boundary of the CT10 uncertainty band, which is based on
a Hessian treatment of the experimental statistical error with
fixed tolerance ($\Delta\chi^2=100$).
The MSTW08 and NNPDF2.1 uncertainty bands are smaller than the
one from the CT10, in particular due to a dynamic and smaller
tolerance ($\Delta\chi^2\sim25$) and a different (Monte Carlo)
sampling of the statistical error and cross validation,
respectively, but also due to different input data, values of
$\alpha_s$, treatments of heavy quarks and experimental systematic
errors, parametrisations etc.
%
The scale uncertainty
(red lines), estimated in the conventional way by varying the
factorization and renormalization scales simultaneously by a factor
of two up and down about the central scale $\sqrt{M_V^2+p_T^2}$,
stays with $\pm10$ to $\pm15$ \% considerably smaller than the 
CT10 PDF uncertainty alone and of course also the envelope of
all three PDF uncertainties. With threshold resummation, computed
e.g.\ with soft-collinear effective theory, the scale uncertainty
reduces to about $\pm5$ \% at large $p_T$  \cite{Becher:2011fc}.
Measurements of electroweak boson
production with transverse momenta of 1 TeV or beyond at
$\sqrt{s}=$ 7, 8 or 14 TeV will thus clearly help to improve
on the determination of the gluon PDF in the large-$x$ regime.

At low $p_T$, corresponding to $x$-values of 0.01 to 0.1, the
situation is quite different. The quark-gluon process still 
dominates, but the gluon PDF is quite well determined here
through the evolution with errors below 10\% at the scale
$Q=M_Z$ (see Sec.\ \ref{sec:2}). At the same time, the up-
and down-quark PDFs are still strongly influenced by the
well-constrained valence contribution. In contrast, the
uncertainty induced by the unphysical scales persists at
the level of 10\% and represents thus the dominant source
of theoretical uncertainty. Taken together, these
observations leave little room for improvement of the gluon
PDF through electroweak boson production at small transverse
momenta.

%
\begin{figure}[!h]
 \centering
  \epsfig{file=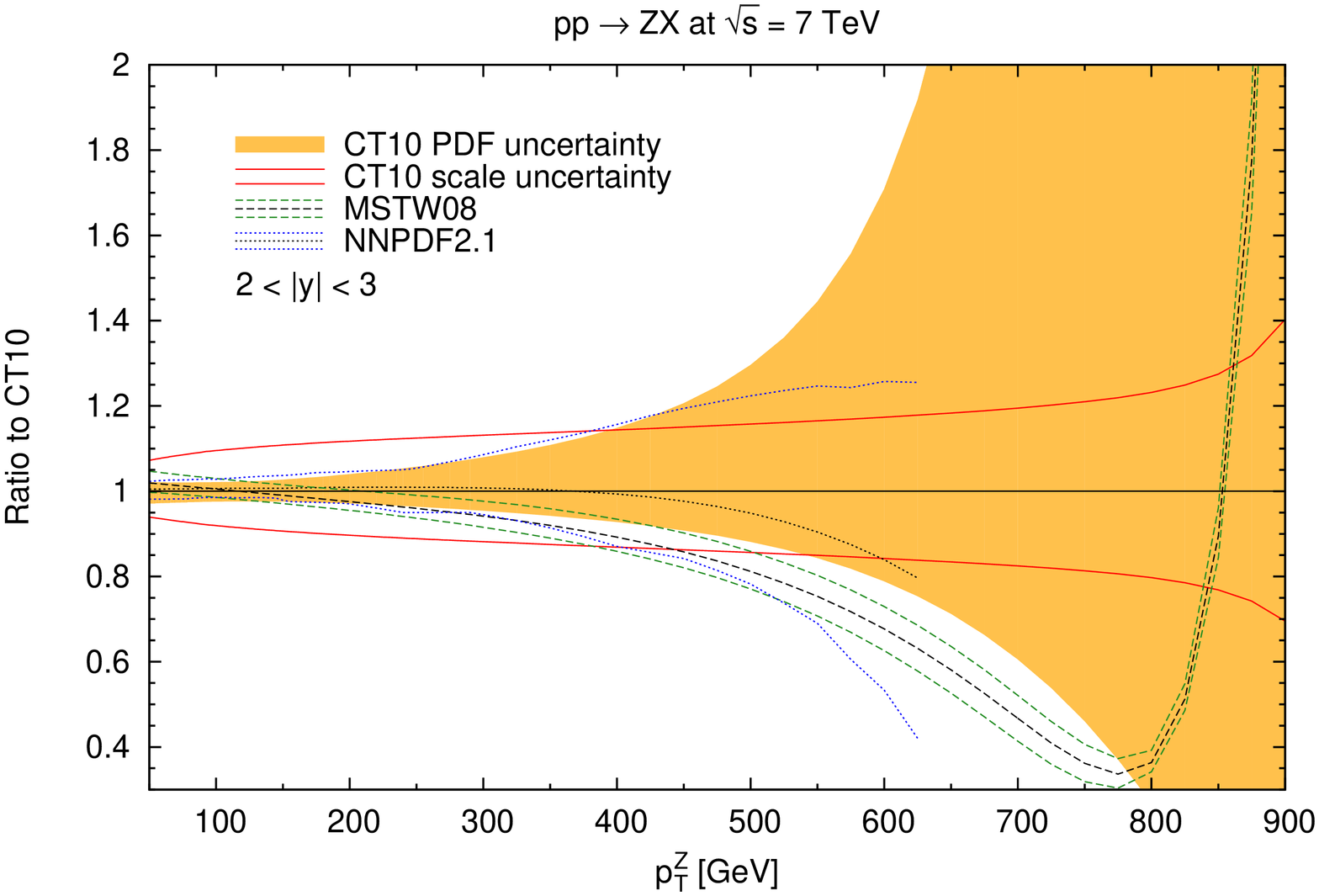,width=0.65\columnwidth,angle=-90}
 \caption{\label{fig:9}Same as Fig.\ \ref{fig:6} for $Z$ bosons at forward rapidity $|y|\in[2;3]$.}
\end{figure}
%

%
\begin{figure}[!h]
 \centering
  \epsfig{file=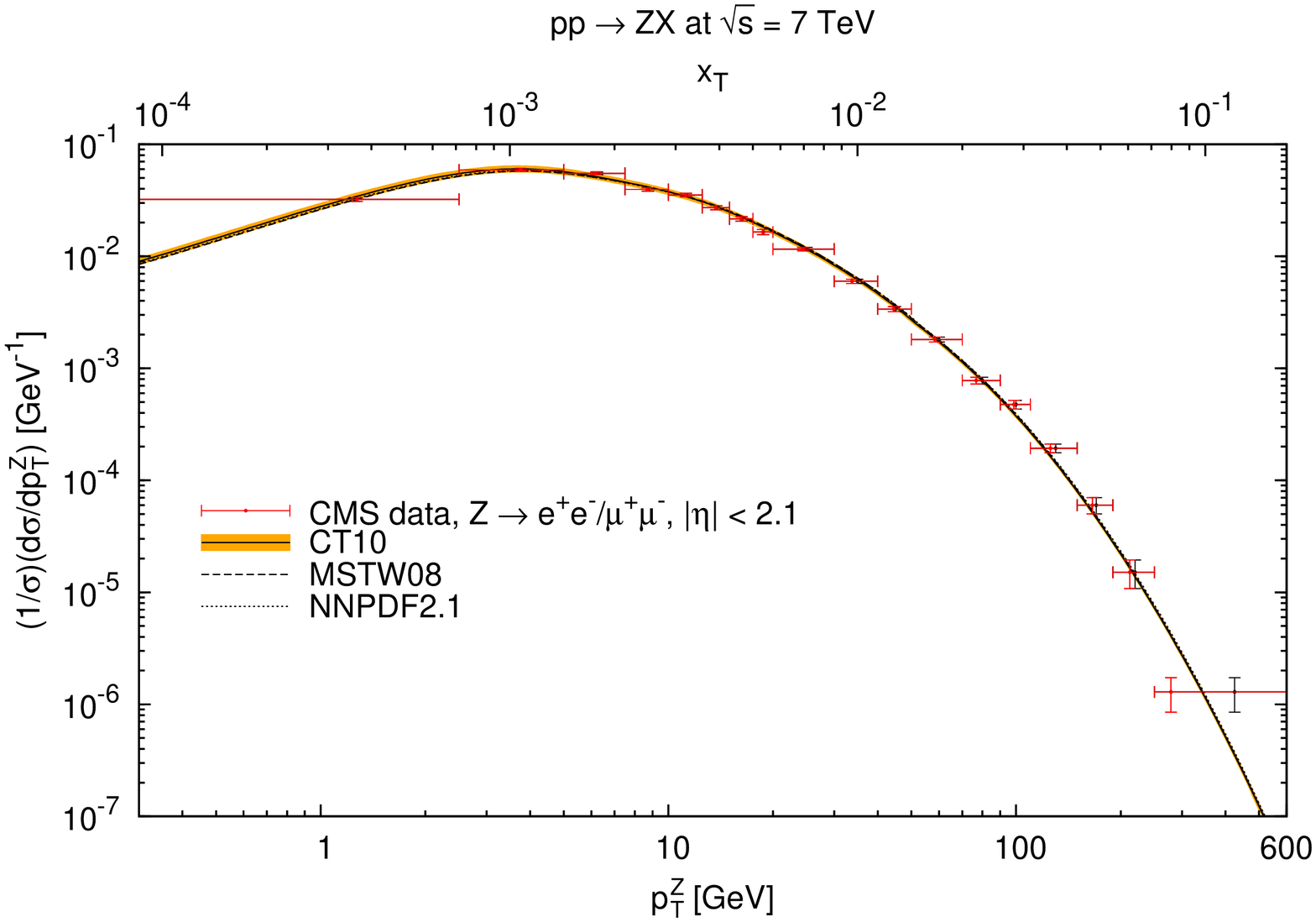,width=0.63\columnwidth,angle=-90}
  \epsfig{file=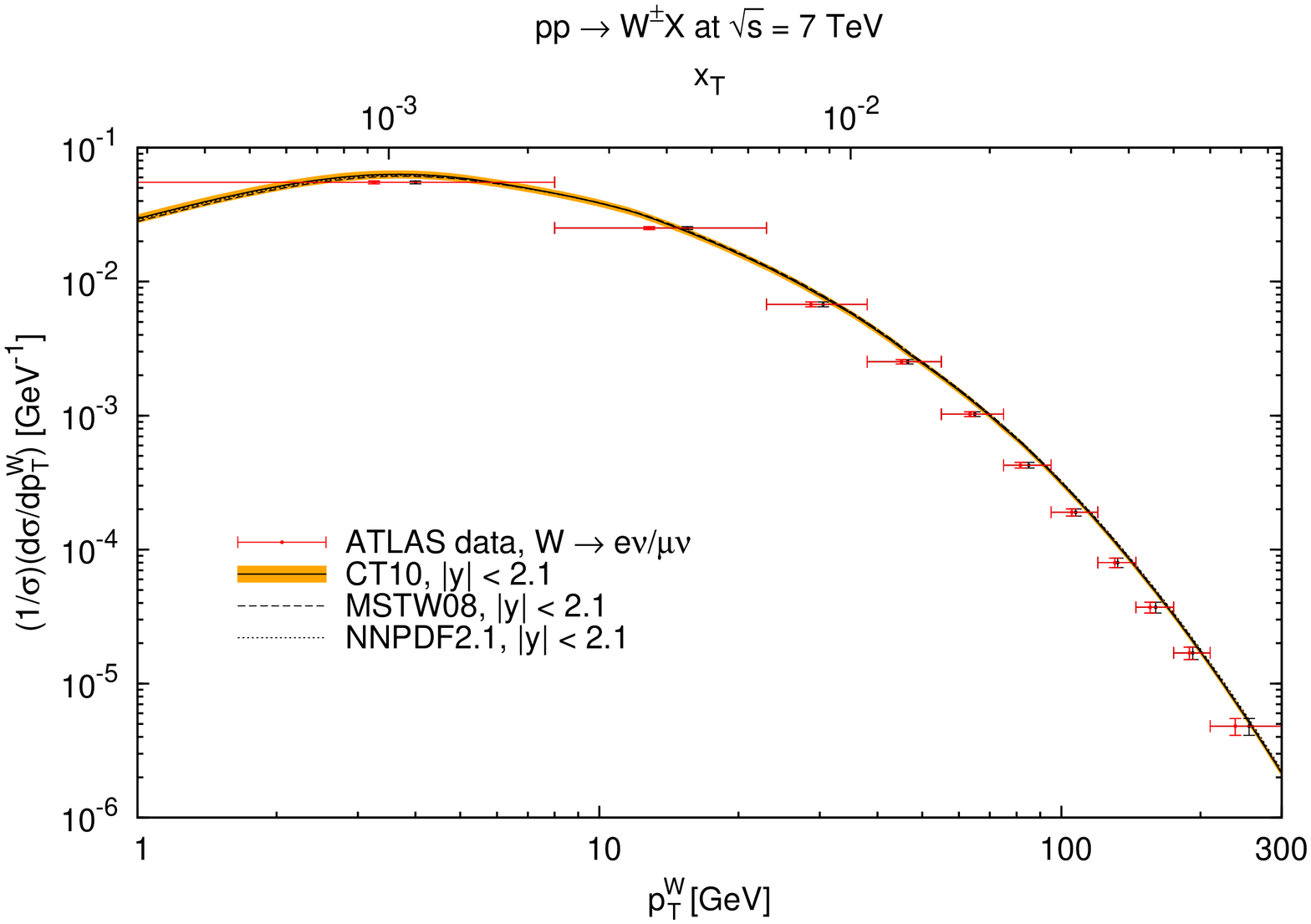,width=0.63\columnwidth,angle=-90}
 \caption{\label{fig:8}Normalized transverse momentum spectra for $Z$ (top) and $W$
 (bottom) bosons. Our theoretical calculations at NLO combined with resummation for
 the rapidity range $|y|<2.1$ and using various PDFs are shown in comparison with CMS
 and ATLAS data on a logarithmic $x$-axis. The data points are positioned at the
 theoretical center of gravity of the bins (red) and at the center of the bins (black).}
\end{figure}
%

Since the rapidity of the produced vector boson enters exponentially
in the expressions for the partonic momentum fractions $x_{1,2}$
(see Sec.\ \ref{sec:2}), it is clear that moving away from central
to forward rapidity creates a more asymmetric situation where the
partons of one incoming proton are probed at much larger, those
of the other proton at much smaller values of $x$. This is reflected in
Fig.\ \ref{fig:9}, where we show cross section ratios obtained with
different PDFs to those obtained with CT10 for the production of electroweak
bosons with rapidities of $|y|\in[2;3]$. These rapidities
are still covered by the CMS and ATLAS electromagnetic endcap
calorimeters, while muons are only detected up to $|\eta|<2.4$ and 2.5,
respectively.
As expected, the PDF uncertainties in the forward region
are much larger and reach easily a factor of two.
For the reasons mentioned above, the MSTW08 uncertainty band is much
smaller than the CT10 band, but has a very different shape, while
the NNPDF2.1 band widens at the same $p_T$-values as the CT10 band,
but can even lead to negative cross sections. The envelope of all
error bands is thus even larger than the error band of CT10 alone.
This demonstrates
the potential of corresponding measurements to pin down the
gluon PDFs, depending on the transverse momenta that can be
reached there.

As a final point, it is also interesting to compare in
Fig.\ \ref{fig:8} in more detail our theoretical predictions
to the experimental data from CMS \cite{Chatrchyan:2011wt}
(top) and ATLAS \cite{Aad:2011fp} (bottom)
on $Z$- and $W$-boson production in the low-$p_T$ regime,
emphasized in this figure by the logarithmic $x$-axis.
The reason is that in this region the theoretical prediction
is also influenced by the parameters $g_1$, $g_2$, and $g_3$
of the non-perturbative function $\tilde{W}^{\rm NP}_{j\bar{k}}
(b,Q,Q_0,x_1,x_2)$ (see Sec.\ \ref{sec:3}), which had been
fitted only to Tevatron run-1 data on $Z$-boson production,
but not yet to LHC data \cite{Landry:2002ix}. As it can be
seen from Fig.\ \ref{fig:8}, this fit allows to describe
also the normalized LHC data perfectly well, so that a newer
fit does not seem necessary or lead to much improvement at
this point. This may, however, change once absolute cross
sections become available.

\section{Conclusions}

In conclusion, we have in this paper investigated a possibility
to better constrain the parton densities in the proton at large
momentum fractions. These parton densities are of fundamental
importance not only for our description of hadronic and nuclear
structure, but also for reliable predictions for new heavy particle
searches at colliders.

After establishing the current status of uncertainty from the
CT10, MSTW08 and NNPDF2.1 parametrizations, we have computed
perturbative and resummed cross sections for electroweak vector boson 
production at the LHC, finding good agreement with published
CMS and ATLAS data for $Z$ and $W$ bosons at $\sqrt{s}=7$ TeV
up to $p_T$ values of 600 and 300 GeV, respectively.
We found that at transverse momenta beyond about 20 GeV, they
were dominated by the QCD ``Compton'' process, inducing a
large sensitivity of the cross sections on the gluon PDFs.

We have shown that with the luminosities reached in the 8
TeV and future 14 TeV runs, transverse momenta in the TeV
range should be measurable, thus providing access to the gluon
PDF at large values of $x$, where it is currently very badly
constrained. The theoretical scale uncertainty has been shown
to stay sufficiently small there.

At smaller transverse momenta, little improvement can be made
on the determination of gluon and quark PDFs through the
proposed process. However, the
uncertainties coming from the resummation calculation and in
particular its non-perturbative component have been shown to
be under control, as new LHC data at $\sqrt{s}=7$ TeV can
be described perfectly well with a fit made only to Tevatron
run-1 data on $Z$-boson production.

We therefore hope that the ATLAS and CMS collaborations will
soon make available analyses of electroweak boson production at
large transverse momenta, so that they can be used in future
global analyses of the parton distribution functions in the proton.
As we have learned, at least the NNPDF collaboration already have
concrete plans to do this \cite{rojo}.

\acknowledgments

We thank M.\ Boonekamp for stimulating our interest on this topic.
This work has been supported by the BMBF Theorie-Verbund ``Begleitende
theoretische Untersuchungen zu den Experimenten an den Gro\ss{}ger\"aten der
Teilchenphysik.''


\end{document}